# The Sun as a Laboratory for Plasma Physics

Arnab Rai Choudhuri
Department of Physics
Indian Institute of Science
Bangalore – 560012

*Several phenomena connected with the magnetic field of the Sun – the cool sunspots, the hot corona, solar flares, the solar wind – are collectively known as solar activity. This paper discusses how one uses the MHD equations to understand how the magnetic field of the Sun is produced by the dynamo process and then gives rise to these diverse activities, making the Sun the best laboratory for plasma physics in the limit of high magnetic Reynolds number (defined at the end of the Introduction).*

## *Introduction*

In elementary textbooks on astrophysics, a star is usually modelled as a spherically symmetric mass of gas bound by its own gravity[1-2]. Our Sun appears much more complicated than such a textbook star. The solar surface is sometimes marked with dark sunspots – their number increasing and decreasing with time with a rough period of about 11 years. Above the solar surface, the solar corona visible during a total solar eclipse has a highly non-symmetric appearance varying with time (i.e. it looks quite different during different eclipses). The corona consists of exceedingly hot gas – its temperature being more than a million degrees in the hottest regions – much higher than the temperature of the solar surface which is close to 6000 K. Sometimes we have extremely violent explosions in the corona known as solar flares. A large flare can disrupt our life in various ways. On 13 March 1989, six million people in Quebec in eastern Canada were without electricity for nearly eight hours due to a blackout caused by a solar flare. All these phenomena which make our Sun much more intriguing than a simple spherically symmetric mass of gas are collectively known as solar activity. Although stars other the Sun appear merely as dots of light even through our largest telescopes, we now have indirect evidence that many stars have activity cycles similar to the Sun – producing starspots much larger than sunspots and stellar flares much more energetic than solar flares[3].

What causes these activities of stars like the Sun is a challenging question in modern astrophysics. A first clue came in 1908 when Hale[4] found evidence of Zeeman splitting in the spectrum of a large sunspot, indicating that sunspots are regions of strong magnetic field of about 0.3 tesla. This was the first discovery of magnetic fields outside the Earth's environment. With this discovery, it became clear that the 11-year sunspot cycle is the magnetic cycle of the Sun. Within the last few decades, astrophysicists have realized that the magnetic field of the Sun is the connecting thread behind all the different manifestations of solar activity[5].

A simple application of the Saha ionization equation suggests that the materials in and around solar-like stars would be in the plasma state, as realized by Saha himself in his very first paper on thermal ionization[6]. Plasmas are known to have complicated interactions with magnetic fields. While studying solar activity, we can usually take the solar plasma to be a continuum. The branch of plasma physics in which we treat the plasma as a continuum fluid is known by the tongue-twisting name magnetohydrodynamics, abbreviated as MHD. We essentially combine the basic equations of fluid mechanics with the basic equations of electromagnetism to obtain the basic equations of MHD[7]. One very important dimensionless number appearing in MHD is the magnetic Reynolds number $R_m = \mu_0 \sigma V L$, where $\mu_0$ is the permeability constant, $\sigma$ is the electrical conductivity, $V$ is the typical velocity and $L$ is the typical dimension of the system. Readers having a knowledge of fluid mechanics will be familiar with Reynolds number defined as $VL/v$, where $v$ is the kinematic viscosity. The magnetic Reynolds number $R_m$ plays a role in MHD somewhat analogous to the role played by Reynolds number in fluid mechanics. Just as a large Reynolds number implies that the effect of viscosity inside the fluid is not important (except in boundary layers), a large magnetic Reynolds number $R_m$ implies that the effect of electrical resistivity inside the plasma is not important. Since $L$ is much larger for astrophysical systems than for laboratory systems, it usually turns out that $R_m << 1$ for laboratory systems and $R_m >> 1$ for astrophysical systems. As a result, magnetic fields behave very differently in astrophysical systems and in laboratory systems.

In the $R_m >> 1$ limit (in which electrical resistivity is unimportant), one can derive a very important result from the basic equations of MHD – often known as Alfven's theorem of flux freezing[8]. According to this theorem, the magnetic field is frozen in a high-$R_m$ plasma and moves with it. The significance of this theorem will be clear when we shall consider its application to the Sun in the next section. In fact, there is a symbiotic relation between solar physics and MHD. The phenomena associated with solar activity require application of MHD for their explanation. On the other hand, some of the most interesting theoretical results of MHD are in the high-$R_m$ limit and cannot be tested in the laboratoraty (where $R_m << 1$). The Sun is a mass of high-$R_m$ plasma much closer to us compared to other astronomical bodies where we may try to look for confirmation of high-$R_m$ MHD results. One often says that the Sun is the best laboratory high-$R_m$ MHD available to us.

### The Sunspots and their Cycle

About a decade after the discovery of magnetic fields in sunspots, Hale et al.[9] made another important discovery. Often two large sunspots appear side by side, as seen in Fig. 1.

Hale et al. found that usually the sunspots in the pair would have opposite magnetic polarities – one being positive and the other negative. The most plausible explanation for this is that there must be a strand of magnetic field underneath the Sun's surface which is poking out through the surface, as shown in Fig. 2(a). If the two intersections of this magnetic strand with the surface become the two sunspots, then we expect one of them to have magnetic field lines coming out so that it becomes a positive-polarity sunspot, whereas the other becomes a negative-polarity sunspot with field lines going down. We shall soon discuss how such magnetic configurations arise. However, if there are such magnetic loops above a pair of sunspots, then the solar atmosphere above the surface should be full of such loops. Now that astronomers have developed techniques for imaging such loops in the solar atmosphere, we know that the solar atmosphere is indeed full of such loops. One beautiful image of coronal loops is shown in Fig. 2(b).

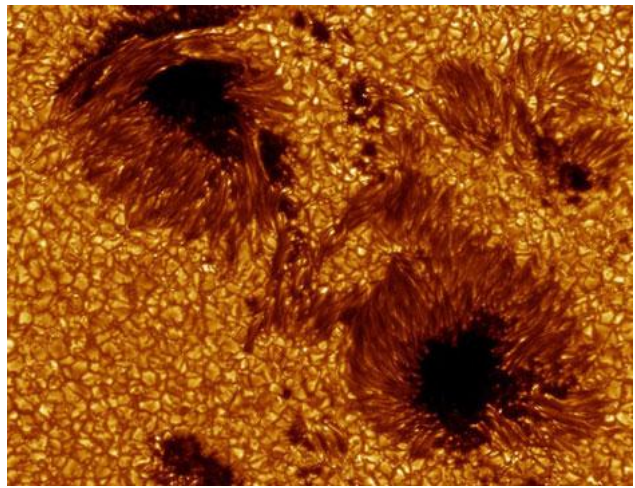

**Fig. 1:** A pair of sunspots on the Sun's surface. Credit: G. Scharmer, K. Langhans and M. Löfdahl, Institute for Solar Physics, Sweden.

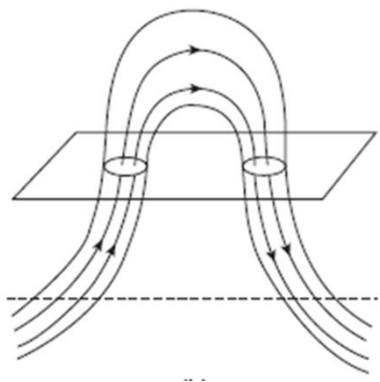 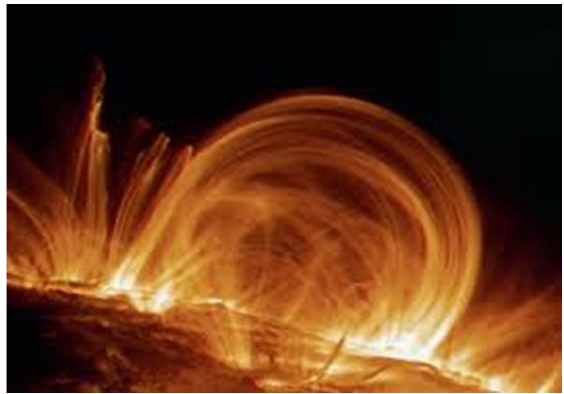

**Fig. 2(a):** A strand of magnetic field underneath the Sun's surface, with a part piercing through the surface to produce a sunspot pair. **Fig. 2(b):** Coronal magnetic loops above the Sun's surface, imaged from the space mission TRACE. Credit: Stanford-Lockheed Institute for Space Research and NASA.

In Fig. 3, we show a magnetogram map of the solar disc. A magnetogram measures the magnetic field at different points on the solar surface, and then such a map is constructed by indicating the regions of positive and negative magnetic polarity respectively with white and black, with grey colour indicating regions where the magnetic field is too weak to measure. In several places, one would see a black patch and a white patch side by side. These would be two sunspots with opposite magnetic polarity. In the northern hemisphere, you see that the positive polarity sunspot (white patch) is to the right of the negative polarity sunspot (black patch), whereas it is the opposite in the southern hemisphere. The magnetic configuration remains like this for a sunspot cycle of 11 years. If you were to look at a magnetogram map during the previous cycle or the next cycle, then we would see black (white) patches to the right of white (black) patches in the northern (southern) hemispheres. In other words, it would take two cycles for the magnetic field to come back to its original configuration.

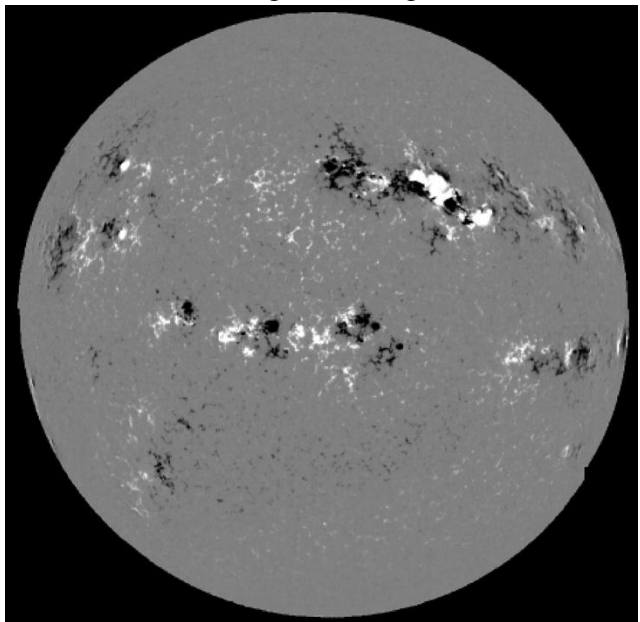

**Fig. 3:** A magnetogram map of the Sun showing the distribution of the magnetic field over the Sun's surface. See the text for explanation.

You notice in Fig. 3 that the two sunspots in a pair are nearly at the same latitude. If such pairs form from magnetic strands underneath the surface as sketched in Fig. 2(a), then these magnetic strands should be from right to left in the northern hemisphere and from left to right in the southern hemisphere. If one introduces a spherical coordinate system with respect to the rotation axis of the Sun, then this subsurface magnetic field would be predominantly in the $\phi$-direction. In the jargon of our subject, a magnetic field with such a configuration is called a toroidal field. On the other hand, a magnetic field with field lines in the poloidal place (such as a dipolar magnetic field) is called a poloidal field.

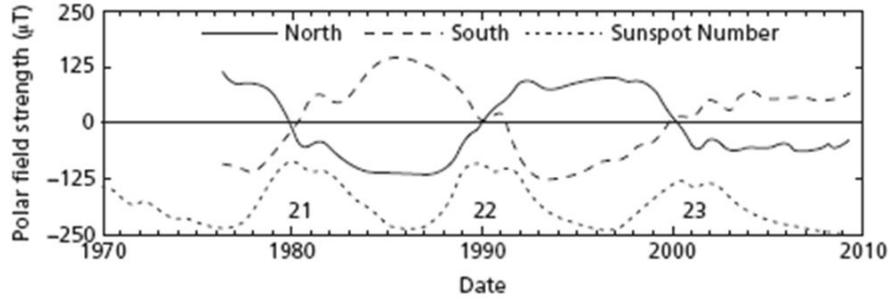

**Fig. 4:** The Sun's polar fields plotted along with the sunspot number starting from the mid-1970s. The two curves in the upper portion show the values of magnetic fields at the two poles (which have opposite signs). The lower dotted curve indicates the sunspot number. Credit: David Hathaway.

In 1955 Parker[10] wrote a paper laying down the foundations of what is called dynamo theory – the theory which explains how magnetic fields arise in astronomical bodies. Parker proposed the bold hypothesis that the sunspot cycle is produced by an oscillation between the toroidal and poloidal magnetic fields – somewhat reminiscent of the oscillation between kinetic and potential energies in a simple harmonic oscillator (although this analogy has many serious limitations). It was in the same year 1955 in which Parker formulated his dynamo theory that the first evidence for a weak magnetic field near the Sun's poles was found[11], indicating that the Sun has a poloidal field. Whether the toroidal and poloidal fields of the Sun really have an oscillation between them as proposed by Parker could be established only several decades later. Fig. 4 shows how the magnetic fields at the two poles of the Sun (measured regularly only from the mid-1970s) varied with time, with the sunspot number plotted below. Remember that sunspots form from the toroidal field and the sunspot number is therefore an indication of the strength of the toroidal field. You can see in Figure 4 that the Sun's polar field (a manifestation of the poloidal field) becomes zero at the time when the sunspot number or the toroidal field is near the maximum. On the other hand, the sunspot number or the toroidal field is close to zero when the polar (i.e. the poloidal) field is the strongest. We thus see an oscillation between the toroidal and the poloidal magnetic fields as envisaged by Parker long before its establishment from observational data.

In order to produce the oscillation between the toroidal and poloidal components as seen in Fig. 4, one needs a mechanism for generating the toroidal field from the poloidal field and another mechanism for generating the poloidal field from the toroidal field. We have already mentioned Alfven's theorem that in high-$R_m$ plasmas magnetic fields are frozen in the plasma and move with it. We can now use this theorem to explain how the toroidal field gets generated from the poloidal field. Fig. 5(a) shows a poloidal magnetic field line. Now, unlike the Earth which rotates almost like a solid body, the Sun is found to have differential rotation. In other words, different regions of the Sun have different angular velocity – regions near the equator having a higher angular velocity compared to regions at higher latitudes. Since magnetic fields would be frozen in the solar plasma, the faster-moving layers near the solar equator would drag the magnetic field line shown in Fig. 5(a) such that after some time it would look as shown in Fig. 5(b). We now see that there is a toroidal part to the magnetic field, although the original magnetic field shown in Fig. 5(a) was completely poloidal. We thus see that the toroidal field can be produced from the poloidal field by the action of differential rotation. We also notice that the toroidal field in the two hemispheres has opposite directions. So, if parts of this toroidal field were to rise to the surface to produce sunspots, then the sunspot pairs in the two hemispheres will have opposite magnetic configurations as seen in Fig. 3.

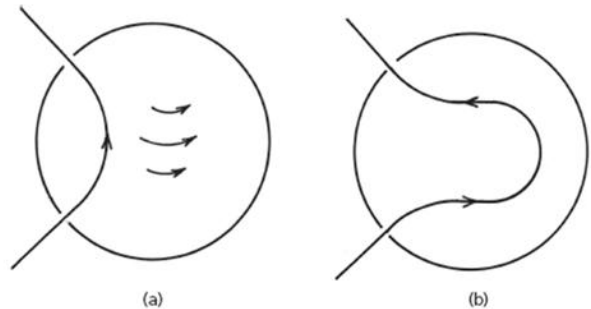

**Fig. 5:** The production of toroidal magnetic field from poloidal magnetic field in the Sun. (a) An initial poloidal field line, with small arrows indicating rotation varying with latitude. (b) A sketch of the field line after it has been stretched by the faster rotation near the equatorial region.

We now come to the question why parts of the toroidal field seen in Fig. 5(b) would float to the surface. Before addressing that question, let us look into the question of why the magnetic field at the solar surface remains bundled up within sunspots rather than filling up all space. It is known that energy is transported outward by convection in the layers of the Sun just underneath its surface. The granular structures that you see in Fig. 1 around sunspots are convection cells. To understand the structure of the magnetic field in this region of convection, we need to study the interaction between the magnetic field and convection – a subject known as magnetoconvection, of which the first linear theory was developed by Chandrasekhar[12]. Since magnetic field lines have tension associated with them, they resist convection. As a result, magnetic fields in convecting plasmas get bundled up in concentrations known as flux tubes[13]. Sunspots are supposed to be such magnetic flux tubes within which energy transport by convection is suppressed by magnetic tension, resulting in the lowered temperature (and the dark appearance) compared to the surroundings.

Let us now consider a part of the toroidal field that may exist in the form of a flux tube. There has to be a pressure balance between the inside and the outside. Since the magnetic pressure is given by $B^2/2\mu_0$, the pressure balance equation is

$$p_{\text{out}} = p_{\text{in}} + \frac{B^2}{2\mu_0},$$

where $p_{\text{out}}$ and $p_{\text{in}}$ are respectively the gas pressure outside and inside the flux tube. It follows from this equation that the pressure inside the flux tube must be less than the pressure outside – which often, though not always, implies that the density inside the flux tube also may be less than the surrounding density. We know from Archimedes's principle that such a system would be buoyant. Parker[14] introduced this concept of magnetic buoyancy, from which it follows that a part of a toroidal field may rise to the surface to produce the sunspot pair, as indicated in Figure 2(a). Our group was one of first groups to carry out detailed simulations of the formation of sunspots by the buoyant rise of flux tubes[15-16].

We have seen above that, starting from the poloidal magnetic field shown in Figure 5(a), we can model various aspects of sunspot pair formation by combining the ideas of flux freezing, magnetoconvenction and magnetic buoyancy. To complete the theory of the solar cycle, we need a mechanism for generating the poloidal field from the toroidal field. How this happens is the central question of solar dynamo theory. It is beyond our scope to discuss this question in this short paper. Some of the other authors (Binod Srinivasan, Kandaswamy Subramanian) in this special issue discuss the basic concepts of dynamo theory. Let us mention only the following here. From the mid-1990s a few type of dynamo model known as the flux transport dynamo model started being developed[17-18]. This model, in the development of which our group at IISc played a leading role, is able to explain many aspects of the solar cycle[19-22].

**Plasma processes in the solar corona**

We have discussed how sunspots form on the solar surface. We have also pointed out that magnetic fields above bipolar sunspot pairs give rise to loops, as shown in Figure 2(b). Now we shall look into the question how the various coronal phenomena connected with magnetic activity arise. The first important question connected with the corona is why it has a temperature much higher than the solar surface – reaching to more than a million degrees in the hottest regions. It is worth mentioning that the thermal ionization theory developed by Saha played a key role in establishing the high temperature of the corona for the first time. Some mysterious spectral lines of the corona were at last identified in 1943 by Edlén[23] as coming from multiply ionized iron atoms (like Fe-XIV). Such a high level of ionization would require a very high temperature of the corona.

We know that a gas heated to temperatures of order millions of degrees emit radiation in X-rays and Extreme UV. Since the upper atmosphere of the Earth is opaque to such radiation, we can image such radiation from an astronomical source only by sending an X-ray or extreme UV telescope in a spacecraft above the Earth's atmosphere. Beginning with Skylab in the 1970s, astronomers have succeeded in obtaining extraordinary images of the solar corona from space. Fig. 6 shows an Extreme UV image of the corona. It is clear that the loops above sunspots are the hottest regions of the corona, making it obvious that the magnetic field arching above sunspots must be playing an important role in producing the high temperature of the corona.

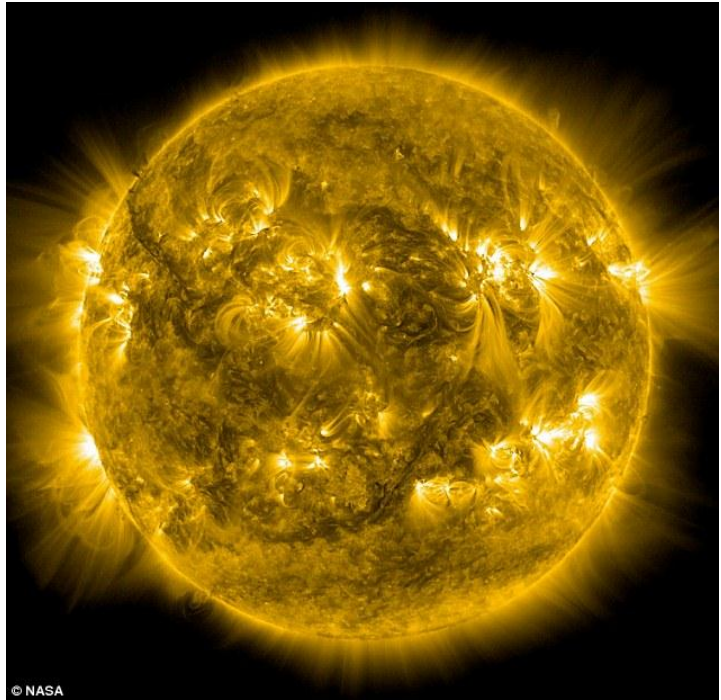

**Fig. 6:** The Sun imaged by the Extreme UV telescope aboard the space mission SOHO. Credit: SOHO (ESA and NASA).

In a provocative paper in 1972, Parker[24] proposed a mechanism by which heat can be generated within the magnetic loops of the corona. Since magnetic field lines of the loops will continue below the solar surface where vigorous convection is going on, the field lines will continuously get disturbed by the underlying convection and will get tangled up within the loop. This would lead to the formation of what are called current sheets – regions where electric currents would be concentrated due to sharp gradients of magnetic field. Heat is expected to be generated within these current sheets inside the loops, causing the high temperature of the loops.

Quite early in 1958 – before the reason behind the high temperature of the corona was known – Parker[25] realized that the high temperature of the corona would have one important consequence. The gravitational field of the Sun will be unable to confine the very hot gas in the corona. As a result, the outer regions of the corona will keep expanding in the form of a plasma flow which Parker called the solar wind. Parker estimated that this solar wind, which will flow through the solar system, will have a velocity of a few hundred km s$^{-1}$ near the Earth's orbit. Within a few years of Parker's prediction, the solar wind was detected by in situ measurements from space missions. We now know that the space surrounding our Earth is not empty, but the Earth is immersed in a plasma flow coming from the Sun – connecting the Earth to the Sun in complicated ways.

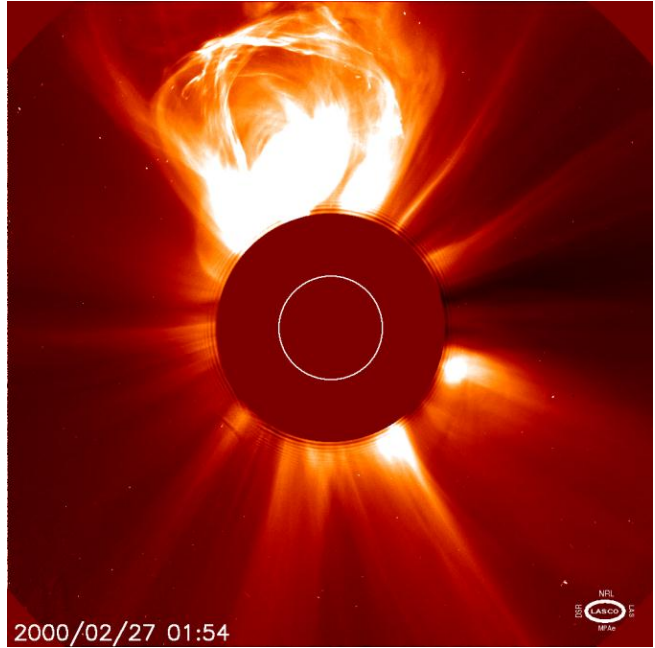

**Fig. 7:** A coronal mass ejection (CME) imaged by the coronagraph LASCO C2 on board the space mission SOHO. Credit: SOHO (ESA and NASA).

There are often gigantic explosions on the Sun – known as solar flares – releasing energy of the order of $10^{32}$ erg. These explosions are believed to be caused by a plasma process known as magnetic reconnection, in which magnetic fields of different orientation on the two sides of a surface are pressed against each other. These explosions often throw up large chunks of plasma from the Sun known as coronal mass ejections (CMEs). Fig. 7 shows a CME leaving the Sun – imaged by a coronagraph in space within which an artificial eclipse is created by blocking the solar disk. These CMEs are eventually carried outward by the solar wind and, if placed in favourable locations, may impinge on the Earth's magnetosphere, thereby giving rise to magnetic storms. Since the solar wind takes about 3-4 days to cover the Sun-Earth distance, that is the usual time after which the effect of a large solar flare is felt on the Earth.

### *Conclusion*

A scientific research field has a special intellectual appeal if the various phenomena in that field can be shown to be connected to each other through a logical chain of arguments. We have tried to give an idea in this paper that the various aspects of solar activity are related to each other through the interconnecting thread of the magnetic field governed by MHD equations. First of all, the magnetic field is produced inside the Sun by the dynamo process. Then it comes out by magnetic buoyancy to produce sunspots on the solar surface. The magnetic loops above sunspots may become full of many current sheets, giving rise to the high temperature of the corona. The hot corona eventually causes the solar wind within which the Earth is immersed. Solar flares taking place due to magnetic reconnection may throw up CMEs which eventually may produce magnetic storms in the Earth. Readers desirous of learning more about the connections among different aspects of solar activity may turn to the advanced textbook by Priest[26].

## *Acknowledgements*

My research is supported by a JC Bose Fellowship awarded by DST.

## *References*

1. D. Maoz, Astrophysics in a Nutshell, Princeton University Press, (2007)
2. A.R. Choudhuri, Astrophysics for Physicists, Cambridge University Press, (2010)
3. A.R. Choudhuri, Starspots, stellar cycles and stellar flares: Lessons from solar dynamo models, *Science China Physics, Mechanics & Astronomy*, **60**, 19601 (2017)
4. G.E. Hale, On the probable existence of a magnetic field in sun-spots, *Astrophysical Journal*, **28**, 315, (1908)
5. A.R. Choudhuri, Nature's Third Cycle: A Story of Sunspots, Oxford University Press (2015)
6. M.N. Saha, Ionisation in the solar chromosphere, *Philosophical Magazine*, **40**, 472, (1920)
7. A.R. Choudhuri, The Physics of Fluids and Plasmas: An Introduction for Astrophysicists, Cambridge University Press, (1998)
8. H. Alfvén, On the existence of electromagnetic-hydrodynamic waves, *Arkiv för matematik, astronomi och fysik*, **29 B**, No. 2 (1942)
9. G.E. Hale, F. Ellerman, S.B. Nicholson, A.H. Joy, A. H., The magnetic polarity of sun-spots, *Astrophysical Journal*, **49**, 153, (1919)
10. E.N. Parker, Hydromagnetic dynamo models, *Astrophysical Journal*, **122**, 293, (1955)
11. H.W. Babcock, H.D. Babcock, The Sun's magnetic field, 1952–1954, *Astrophysical Journal*, **121**, 349, (1955).
12. S. Chandrasekhar, On the inhibition of convection by a magnetic field, *Philosophical Magazine*, **43**, 501, (1952).
13. N.O. Weiss, The expulsion of magnetic flux by eddies, *Proceedings of the Royal Society of London. Series A*, **293**, 310, (1966)
14. E.N. Parker, The formation of sunspots from the solar toroidal field, *Astrophysical Journal*, **121**, 491, (1955)
15. A.R. Choudhuri, P.A. Gilman, The influence of the Coriolis force on flux tubes rising through the solar convection zone, *Astrophysical Journal*, **316**, 788, (1987)
16. S. D'Silvs, A.R. Choudhuri, A. R., A theoretical model for tilts of bipolar magnetic regions, *Astronomy and Astrophysics*, **272**, 621, (1993)
17. A.R. Choudhuri, M. Schüssler, M. Dikpati, The solar dynamo with meridional circulation, *Astronomy and Astrophysics*, **303**, L29 (1995)

18. B.R. Durney, On a Babcock–Leighton dynamo model with a deep-seated generating layer for the toroidal magnetic field, *Solar Physics*, **160**, 213, (1995)

19. A.R. Choudhuri, The origin of the solar magnetic cycle, *Pramana*, **77**, 77, (2011)

20. A.R. Choudhuri, The irregularities of the sunspot cycle and their theoretical modeling, *Indian Journal of Physics*, **88**, 877, (2014)

21. P. Charbonneau, Solar dynamo theory, *Annual Review of Astronomy and Astrophysics*, **52**, 251, (2014)

22. B.B. Karak, J. Jiang, M.S. Miesch, P. Charbonneau, A.R. Choudhuri, Flux transport dynamos: From kinematics to dynamics, *Space Science Reviews*, **186**, 561, (2014)

23. B. Edlén, Die Deutung der Emissionslinien im Spektrum der Sonnenkorona, *Zeitschrift für Astrophysik*, **22**, 30, (1943)

24. E.N. Parker, Topological dissipation and the small-scale fields in turbulent gases, *Astrophysical Journal*, **174**, 499, (1972)

25. E.N. Parker, Dynamics of the interplanetary gas and magnetic fields, *Astrophysical Journal*, **128**, 664, (1958)

26. E.R. Priest, Magnetohydrodynamics of the Sun, Cambridge University Press, (2014)